\documentclass[prl,twocolumn,showpacs,preprintnumbers,floatfix,superscriptaddress,10pt]{revtex4}

\newcommand{\be}{\begin{equation}}
\newcommand{\ee}{\end{equation}}
\newcommand{\ba}{\begin{eqnarray}}
\newcommand{\ea}{\end{eqnarray}}

\begin{document}

\title{Mutual friction in a cold color flavor locked superfluid and r-mode instabilities in compact stars}

\author{Massimo Mannarelli}
\affiliation{Instituto de Ciencias del Espacio (IEEC/CSIC) Campus Universitat Aut\`onoma de Barcelona, Facultat de Ci\`encies, Torre C5, E-08193 Bellaterra (Barcelona), Spain}

\author{Cristina Manuel}
\affiliation{Instituto de Ciencias del Espacio (IEEC/CSIC) Campus Universitat Aut\`onoma de Barcelona, Facultat de Ci\`encies, Torre C5, E-08193 Bellaterra (Barcelona), Spain}

\author{Basil A. Sa'd}
\affiliation{%
Frankfurt Institute for Advanced Studies,
J.W. Goethe-Universit\"{a}t,
D-60438 Frankfurt am Main, Germany}%
\affiliation{Institut f\"ur Theoretische Physik,
J.W.\ Goethe-Universit\"at,
D-60438 Frankfurt am Main, Germany}

\begin{abstract}
Dissipative processes acting in  rotating neutron stars are essential in preventing the growth of the r-mode instability.  We estimate the damping time of r-modes of an hypothetical compact quark star made up by color flavor locked quark matter at a temperature $T \lesssim 0.01$ MeV. The dissipation  that we consider is  due to the  the mutual friction force between the normal and the superfluid component  arising from the elastic scattering of phonons with quantized  vortices. This process is the dominant one for   temperatures $T \lesssim 0.01$ MeV where the mean free path of phonons due to their self-interactions is larger than the radius of the star.   We find that r-modes oscillations are  efficiently damped by this mechanism for pulsars rotating at frequencies of the order of $1$ Hz at most. Our analysis rules out the possibility that   cold pulsars  rotating at higher frequencies  are entirely made up by color flavor locked quark matter.
\end{abstract}

\date{\today}
\maketitle

{\it Introduction.---} According to Quantum Chromodynamics (QCD), at asymptotically high densities strong interacting matter cannot be described as made up of nucleons; at sufficiently high densities nucleons are crushed  and their quark matter content should be deconfined~\cite{Collins}. It remains an open question whether deconfined quark matter can actually  exist in neutron stars that have central density at most $\sim 10$ times that of ordinary nuclear matter or whether stars entirely consisting of quark matter can exist. The latter case may be realized if strange  quark matter is absolutely stable with a lower energy per baryon than nuclear matter.  Such a quark star  would  consist of quark matter fluid all the way to the surface. 

If deconfined quark matter exists in compact stars  it is likely to be in one of the possible color superconducting phases~\cite{reviews}. Neutron stars that are more than  a few seconds old have temperatures  of the order of tens of keV at most and  such temperatures are smaller than the  critical temperatures of color superconductors  which are generically of order tens of MeV~\cite{reviews}. The color superconducting phase that is energetically favored at extremely high densities and low temperatures   is the color flavor locked  phase  (CFL)~\cite{Alford:1998mk}. In this phase $u$, $d$ and $s$ quarks pair forming a quark condensate that is antisymmetric in color and flavor indices. The  order parameter breaks the baryonic number $U(1)_B$ symmetry spontaneously, and therefore CFL quark matter is  a superfluid.

In this Letter we analyze the r-mode instability of compact stars entirely made up of CFL quark matter.
R-modes are non-radial oscillations of the star with the Coriolis force acting as the restoring force, they provide a severe limitation on the star's rotation frequency through coupling to  gravitational radiation (GR) emission~\cite{Andersson:2000mf}. When dissipative phenomena damp these r-modes the star can rotate without losing  angular momentum to GR. If dissipative phenomena are not strong enough,  these  oscillations  
will grow exponentially and the star will keep slowing down until some dissipation mechanism   can  damp the r-modes. Therefore,  the study of r-modes  is useful in  constraining  the stellar structure and can be used to rule out some color superconducting phases~\cite{Madsen:1999ci}.

 We  consider only the low temperature regime ($T \lesssim 0.01$ MeV)  where the thermal and dynamical properties of the system are dominated by the contribution of the superfluid phonons $\varphi$ associated with the spontaneous breaking of $U(1)_B$. In order to provide an efficient mechanism for damping of r-modes in Ref.~\cite{Manuel:2004iv}  the processes $\varphi \leftrightarrow \varphi \varphi$ and  $\varphi \varphi \leftrightarrow \varphi \varphi$ were studied.
The latter process is largely suppressed, while the former can be seen to produce a  mean free path that scales in terms of the quark chemical potential, $\mu$, and temperature as $\ell \sim \frac{\mu^4}{T^5} $. For low temperatures the authors of~\cite{Manuel:2004iv} find that $\ell$ exceeds the radius of the neutron  star (of order $10$ Km)  and therefore this process does not provide an efficient dissipative mechanism and cannot be responsible for the damping of r-modes.  Actually in this temperature regime one can treat the phonon system as an ideal bosonic gas. 

In the present paper we study a different dissipative mechanism which  operates in rotating superfluids due to the presence of quantized vortices. In this case a mutual friction between the superfluid and the normal component arises because of the scattering of phonons off superfluid vortices. 
We estimate the time scale for the dissipation of energy associated with this process and conclude that for low temperatures and for reasonable values of the vortex mass this mechanism is  efficient in damping r-mode instabilities for stars rotating at frequencies $\nu \lesssim 1$ Hz. On the other hand, cold pulsars rotating at higher frequencies are unlikely  to be CFL quark stars. As a matter of fact in the  temperature range considered here the populations  of electrons~\cite{Rajagopal:2000ff} and  gapped quarks~\cite{Madsen:1999ci,Sa'd:2008gf}   are  exponentially suppressed and they cannot significantly contribute to dissipative processes.

We leave for a future project the study of the effect of mutual friction in the dissipation of r-modes in a higher $T$ regime of the CFL phase. An incomplete analysis of the damping mechanisms as arising from other dissipative channels has been carried out in Refs.~\cite{Madsen:1999ci,Sa'd:2008gf}.

Throughout, we use natural units $\hbar=c =K_B =1$ and Minkowski metric $\eta_{\mu\nu} = (1,-1,-1,-1)$.

{\it Vortex dynamics.---}  We review the description of the forces acting  on a vortex line pioneered by Hall and Vinen~\cite{HallV,Sonin:1987zz}.  

The relative motion of the superfluid component with respect to the vortex produces a Magnus force per unit vortex length given by 
\begin{equation}
{\bf F}^M=\kappa \rho_s ({\bf v}_s - {\bf v}_L) \times \hat {\bf z} \,,
\label {magnus}
\end{equation}
where $\kappa$ is the circulation, $\hat {\bf z}$ indicates the direction of the vortex line, $\rho_s$ is the superfluid density,  ${\bf v}_s$ is the superfluid (local) velocity and ${\bf v}_L$ is the vortex velocity. 

The normal component reacts to the vortex motion through a  force which can be decomposed into components parallel and perpendicular to the relative motion ${\bf v}_n - {\bf v}_L$,
\begin{equation}
{\bf F}^N=D({\bf v}_n -{\bf v}_L)+D' \hat {\bf  z} \times ({\bf v}_n - {\bf v}_L) \,,
\label{frictionforce}
\end{equation}
where ${\bf v}_n$ is the  velocity of the normal component and $D$ and $D'$ are two coefficients  that depend on the microscopic physics of the system. At  low temperature, these coefficients  can be computed  by analyzing the scattering of quasiparticles off a vortex~\cite{Sonin:1987zz}. 

At equilibrium  these two forces  add up to zero
\begin{equation}\label{equilibrium}
{\bf F}^N + {\bf F}^M = 0,
\end{equation}
and from this equation one can determine the vortex velocity  to be given by
 \begin{equation}
{\bf v}_L = {\bf v}_s + \alpha'({\bf v}_n - {\bf v}_s) + \alpha \hat {\bf z} \times ({\bf v}_n - {\bf v}_s),
\label{vl}
\end{equation}
where 
\begin{equation}
\alpha = \frac{d_{||}}{d^{2}_{||}+(1-d_\bot)^2}\ , \qquad 1-\alpha'=\frac{1-d_{\bot}}{d_{||}^2+(1-d_\bot)^2},
\end{equation}
where
$d_{||}=D/\kappa \rho_s$ , and  $d_{\bot}=D'/\kappa \rho_s$.

Since ${\bf F}^M$ is the force  acting on the vortex generated by the superfluid component,  the corresponding  force on the superfluid   is given by $-{\bf F}^M$ and we will assume that there are no other forces acting on the superflow.

Upon substituting Eq.~(\ref{vl}) in Eq.~(\ref{magnus}) one finds that the mutual friction force on a unit volume of the superfluid component per unit vortex length  is given by,
\begin{equation}
 {\bf F}^{SN}=-N_v \kappa \rho_s  \alpha({\bf v}_s - {\bf v}_n) + N_v \kappa \rho_s \alpha' \lbrack \hat {\bf z} \times ({\bf v}_s - {\bf v}_n)\rbrack \,,
\label{mff}
 \end{equation}
where we have multiplied the force by the surface density of vortices $N_v$. This force is acting  on the superfluid component due to the interaction of the normal fluid component with the vortices. The first term on the right hand side is responsible for dissipation.

The expression in Eq.~(\ref{mff}) is valid if the Magnus force and the mutual friction force equilibrate, i.e. if Eq.~(\ref{equilibrium}) is satisfied. If we introduce a perturbation of the superfluid velocity it is not guaranteed that these two forces will equilibrate and we will have that the force per unit vortex length acting on the vortex is given by
\be\label{offequilibrium}
\delta {\bf F}^v \equiv \delta{\bf F}^N +\delta {\bf F}^M = m_v \frac{d \delta{\bf v}_L}{d t}\, ,
\ee
where $\delta {\bf v}_L$ is the variation of the vortex velocity induced by the perturbation of the superfluid velocity and $m_v$ is the mass per unit length of the vortex (we will loosely refer to this as the vortex mass). 

We will assume that when an  external  perturbation is applied to the system the normal component    remains in thermal equilibrium. On the other hand, the superfluid and vortex velocities will have an oscillatory behavior
\begin{equation}
\delta {\bf v}_{(s,L)} = \delta {\bf v}_{(s,L)(0)} e^{i \omega t}\,,
\end{equation}
where $\omega$ is the frequency of the r-mode perturbation, which is of the order of the rotation frequency of the star \cite{Andersson:2000mf}. Henceforth we shall drop out the ``$(0)$'' from the subscripts. We find that the force acting on a unit volume of the superfluid component becomes,
\begin{equation}
 \delta {\bf F}^{SN} = N_v \kappa \rho_s (-\tilde \alpha \delta {\bf v}_s +\tilde \alpha'\hat {\bf z} \times \delta {\bf v}_s) \,,
\end{equation}
where $\tilde \alpha', ~ \tilde \alpha$ are defined as $\alpha',~\alpha$ with the replacement $ d_{||} \rightarrow \tilde d_{||} \equiv d_{||}+i\left( \frac{\omega m_v}{\kappa \rho_s}\right)$.

{\it Phonon-vortex scattering.---} The rotating CFL superfluid  is threaded with vortex lines whose density per unit area is $N_v = \frac{2 \Omega}{\kappa}$, where $\Omega$ is the rotation frequency of the star and where the quantized circulation 
is $\kappa=  \frac{2\pi}{\mu}$~\cite{Iida:2002ev}.

In order to obtain the coefficients $D$ and $D'$ 
 appearing in the friction force in Eq.~(\ref{frictionforce}) for the cold CFL matter
one has to evaluate the cross section associated with  the  scattering of phonons
with vortices.  We will restrict our analysis to the elastic scattering because inelastic processes are suppressed, see~\cite{Fetter}, and   consider  phonons as a  bosonic ideal gas. In this case one can describe the problem of  phonon-vortex interactions using  gravity analogue models~\cite{Unruh,Volovik:1998pc,Mannarelli:2008jq}.

The action of a phonon field moving on the top of a superfluid background is the same as that of  a boson propagating in a given curved space-time. One can consider the superfluid phonon as propagating   along the geodesics of the metric
\be
ds^2 =  {\cal G}_{\mu \nu} dx^\mu d x^\nu \,,
\ee
where ${\cal G}_{\mu \nu}$ is the relativistic acoustic metric~\cite{Mannarelli:2008jq}
\begin{equation}
\label{phonon-metric}
{\cal G}_{\mu \nu} =   
\eta_{\mu\nu} + \left(c_s^2- 1 \right)  v_\mu  v_\nu \,,
\end{equation}
which depends on the speed of sound $c_s$ and the superfluid velocity $v_\mu$. 
In cylindrical coordinates $(t,r,\theta,z)$ far away from the vortex one has that $v^\mu \simeq (1,0, \frac{\kappa}{2 \pi r},0)$  and one can derive   the cross section for the phonon-vortex scattering from the equation of motion of the phonon field in the corresponding curved background. For a phonon of energy $E$  the differential cross section per unit vortex length for the phonon-vortex scattering is given by 
\be
\frac{d\sigma}{d\theta} = \frac{c_s}{2 \pi E} \frac{\cos^2{\theta}}{\tan^2{\frac\theta 2}} \sin^2{\frac{\pi E}{\Lambda}} \,,
\ee
where 
$1/\Lambda =\left(1- c_s^2\right)\kappa/(2 \pi c_s^2)$.
Notice that for the CFL superfluid 
there is an extra factor $(1-c_s^2)$ in the expression of $1/\Lambda$ with respect to the result of~\cite{Sonin:1987zz,Volovik:1998pc} due to the fact 
that the speed of sound is relativistic.

The parallel and transverse cross sections for the phonon-vortex interaction, for   $E \ll \Lambda$, are given by 
\be
\sigma_{\parallel} = \left(1-c_s^2\right)^2
\frac{\kappa^2}{8  c_s^2} k\,, \qquad 
\sigma_{\perp}  = \left(1-c_s^2\right)\frac{\kappa}{c_s} \,,\label{cross-section}
\ee
where  $k$ is the phonon momentum.  A similar  result has been derived  for  non-relativistic superfluids by Volovik~\cite{Volovik:1998pc} employing the analogy with cosmic strings in (3+1)d and  also with the problem of a spinning particle in (2+1)d gravity. However, in order to derive the exact  phonon-vortex cross section one has to consider  corrections due to the short distance behavior of the velocity field~\cite{Volovik:1998pc}.

From the expressions of the cross sections in Eqs.~(\ref{cross-section})   one can determine the coefficients $D$ and $D^\prime$ that appear in the mutual friction force (see e.g.~\cite{Sonin:1987zz})
\be
D= \sigma_{\parallel} c_s \rho_n \,,\qquad
D^\prime = \sigma_{\perp} c_s \rho_n \,,
\ee
where $\rho_n$ is the phonon density.

{\it Mutual friction time scale.--} We now discuss the dissipation of r-modes due to the mutual friction force in the CFL phase at small temperatures. In the reference frame where the normal component velocity is zero
the friction force due to  a superfluid velocity perturbation is 
 \begin{equation}
 \delta {\bf F}_{diss}^{SN}=-N_v\, \kappa\, \rho_s\, \tilde\alpha\, \delta {\bf v}_s \,,
 \end{equation}
and the corresponding  energy density dissipation is 
 \be
 \left( \frac{d \cal E}{dt} \right)_{MF} = \delta  {\bf F}_{diss}^{SN} \cdot \delta {\bf v}_s^* 
 = -N_v \kappa \rho_s \tilde\alpha (\delta {\bf v}_s \cdot \delta {\bf v}_s^*).
 \ee
The typical dissipation time is given by 
\begin{equation}
 \frac{1}{\tau_{MF}} = -\frac{1}{2\cal E} \left( \frac{d \cal E}{dt} \right) _{MF} =2 \Omega  Re (\tilde\alpha) \,,
 \end{equation}
where we used the fact that  $ {\cal E} = \frac{1}{2} \rho_s (\delta {\bf v}_s \cdot \delta {\bf v}_s^*)$. 
Since the expression  of $\tilde\alpha$ depends on the ratio $x=\frac{\omega m_v}{\kappa \rho_s}$ we will consider different values of this quantity.

In case of  negligible values of the vortex mass, i.e. $x \ll 1$, we have that 
\begin{equation} 
\tilde\alpha \simeq \alpha\simeq \left(1-c_s^2\right)^2\frac{2 \pi^5}{405 c_s^6} \left( \frac{T}{\mu}\right)^5\, ,
\end{equation}
and taking $c_s=1/\sqrt{3}$ we obtain
 \begin{equation}
	\frac{1}{\tau_{MF}}= 2 \alpha \Omega \simeq 2 \times 18.1 \left(\frac{T}{\mu}\right)^5 \Omega \,.
 \end{equation}
 Since $T \ll \mu$ the dissipation time scale in this case is much longer than the period of rotation of the star. If we take the opposite limit, $x \gg 1$,  we have that $\frac{1}{\tau_{MF}} \sim \alpha \Omega/x^2$ and the corresponding time scale is  even longer. 
The only case where dissipation time scale is of the same order or less than the period of rotation of the star  is for $x\sim 1$.  
Estimates of the value  of the vortex mass give $m_v \sim \rho\, \xi^2$, where $\xi$ is the radius of the vortex core. The value of $\xi$ is uncertain~\cite{Randeria}, but considering the largest possible estimate where $\xi \sim 1/\Delta$, with $\Delta \gtrsim 10 $ MeV  the  superconducting gap, we obtain that $x \lesssim 10^{-17}$ and then  the effect of the vortex mass should  be negligible.

{\it Critical frequency---}  Now that we have evaluated  the dissipative timescale for the mutual friction, we can calculate the critical frequency above which the r-mode instability appears. 

 We need to compare the timescale for the mutual friction with the ``growth'' timescale for the r-modes $\tau_{GR}$. For a quark star with a  polytropic equation of state with  index $n=1$ one has that~\cite{Andersson:2000mf}
\begin{equation}
  \frac{1}{\tau_{GR}}=\frac{1}{3.26}\left( \frac{\Omega^2}{\pi G_N \rho} \right)^3 \,.
\end{equation}

 The condition of stability is $\frac{1}{\tau_{MF}}>\frac{1}{\tau_{GR}}$, which for $x \ll1$ translates into
\begin{equation} 
\nu \leq \nu_c \simeq 2\times 10^5\left( \frac{T}{\mu} 
\right) {\rm Hz}\,,\end{equation}
 for a canonical star with $M=1.4 M_{\odot}$ and $R=12$ Km. $G_N= 6.6742 \times10^{-8} \,\mbox{cm}^3 \mbox{g}^{-1} \mbox{s}^{-2}$ is the gravitational constant, and $\nu_c$ is the critical rotation frequency. For $T=0.01$ MeV, $\mu = 400$ MeV, we get $\nu_c \simeq 1$ Hz.
 In the unlikely case that the vortex mass is such that $x\sim 1$, we obtain  $\nu_c \simeq 10^8$ Hz.

{\it Discussion.---} Mutual friction in the cold CFL superfluid can damp the  r-modes   if the   rotation frequency of the  star is less or of the order of $1$ Hz. R-modes  in cold CFL stars rotating at larger frequencies cannot be damped by this mechanism.

We have obtained this result studying the interaction  arising from the elastic
scattering of phonons off superfluid vortices.  
In the regime $T \lesssim 0.01$ MeV  mutual friction forces  due to other quasiparticles   present in CFL quark matter   are exponentially suppressed and  do not contribute significantly in damping r-modes oscillations.

Mutual friction in conventional neutron stars has been considered  in Refs.~\cite{Lindblom:1999wi,Andersson:2005rs}. The analysis in that case is more complicated due to the existence of both neutron and proton superfluids and an entrainment effect between them.  The relevant microscopic process is  the scattering of electrons off the magnetic
fields entrapped in the core of the superfluid neutron vortices. In that case, the associated
damping times depend on the poorly known entrainment parameter. In some range of
parameters the resulting  forces are strong enough to damp effectively the r-modes.

Let us remark that  a different possibility not ruled out  by the present analysis is that cold pulsars rotating at frequencies larger than $\nu_c$ are  stars that have only a CFL core and   the r-mode instability is damped by processes arising in  matter constituting the remaining part of the star.  

The majority of  observed  pulsars  have considerably larger frequency than $1$ Hz \cite{Manchester:2004bp},  which indicates that it is unlikely that these pulsars are  entirely made of CFL quark matter.
 Less than $25\%$ of the observed pulsars have rotation frequencies less than $1$ Hz \cite{Manchester:2004bp}; being isolated pulsars, their masses are unknown.  Other observables should
be studied in order to see whether these slowly rotating stars are potential candidates for containing CFL matter.

An alternative picture is that cold rapidly rotating  pulsars are quark stars  which do not consist of  CFL quark matter. When color neutrality and the effect of the strange quark mass are taken into account~\cite{reviews} a promising candidate is the three-flavor crystalline color superconducting (CSC) phase~\cite{Casalbuoni:2005zp}. This color superconducting phase  has   gapless quarks  as well as electrons that can provide r-mode damping. Moreover  CSC quark matter, differently from CFL quark matter, has a large shear modulus~\cite{Mannarelli:2007bs} which is a necessary ingredient in explaining pulsar glitches. Another possibility is that the CFL-$K^0$ phase is realized~\cite{Bedaque:2001je}, however the contribution of kaons to dissipative processes becomes relevant for temperatures $T \gtrsim 1 $ MeV~\cite{Alford:2008pb}. It would be interesting to study mutual friction in the cold regime of this phase because of the presence of other light degrees of freedom.

{\bf Acknowledgments:}
We thank Igor Shovkovy for useful discussions.
This work has been supported by the Spanish grants
AYA 2005-08013-C03-02 and FPA2007-60275.

\end{document}